# UJI PERFORMA DAN *WEBSITE RESPONSIVENESS* INSTITUSI DAN *SMART CITY* SE-JAWA BARAT

**Muhammad Anis Al Hilmi** [1], **Muhamad Mustamiin** [2], **Achmad Nagi** [3], **Adi Suheryadi** [4], **dan Fachrul Pralienka Bani Muhammad** [5]

[1,2,4,5] D3 Jurusan Teknik Informatika, Polindra, Jalan Lohbener Lama No. 8 Indramayu, 45252
[3] Universitas 17 Agustus 1945 Cirebon, Jalan Perjuangan No. 17 Cirebon, 45132

E-mail: alhilmi@polindra.ac.id

**Abstract**

The Government through the Ministry of Information and Communication Technology (Kemkominfo) held the 100 Smart Cities Program, an effort to improve the quality of public services. The program produces a Smart City Masterplan to develop regencies and cities. Access to the website of the regencies-cities government that is member of the program is one success indicator of using technology for the community, because the accessing speed (loading time) and responsiveness of the website is a convenience factor in service users. Hence, the importance of viewing and monitoring public service websites related to public service information in loading time and responsiveness of the website. From the data of regencies-cities registered in this program, many of them are from West Java Province, recorded around 11 of the total 26 regencies-cities. This study presents a website assessment using weighting method that focuses on loading time and responsiveness to see the performance of community services provided by the municipal governments that are incorporated in a smart city case study in West Java Province. There are six parameters in weighting performance of website used in this study, namely first-contentful-paint, first-meaningful-paint, speed-index, interactive, first-cpu-idle, and max potential first input delay. The weighting value is compared with Google's dataset, CrUX (Chrome User Experience Report). The test results show that the majority of websites have an average performance of 38.7 (mobile) 63.6 (web) of the 530 websites incorporated in the smart city, with the best results obtained by the City of Bandung.

**Abstrak**

Pemerintah melalui Kementrian Teknologi Informasi dan Komunikasi mengadakan Program 100 Kota-Kabupaten *Smart City,* sebagai upaya peningkatan kualitas pelayanan publik tersebut. Program tersebut menghasilkan *Masterplan Smart City* yang tujuannya untuk mengembangkan kota-kabupaten. Performansi akses menuju *website* pemerintah kota-kabupaten yang tergabung pada program *smart city* menjadi salah satu indikator keberhasilan pemanfaatan teknologi bagi masyarakat, karena kecepatan mengakses dan responsifitas *website* merupakan faktor kemudahan dalam pengguna layanan. Oleh karenanya, pentingnya melihat dan memonitoring *website* pelayanan masyarakat terkait informasi layanan publik dalam akses waktu tunggu maupun responsivitas dari *website* tersebut. Dari data kabupaten/kota yang terdaftar dalam program ini, banyak di antaranya berasal dari Provinsi Jawa Barat, tercatat sekitar 11 dari total 26 kabupaten/kota. Penelitian ini menyajikan penilaian *website* menggunakan pembobotan yang berfokus pada waktu tunggu dan responsivitas untuk melihat performansi layanan masyarakat yang disediakan oleh pemerintah kota-kabupaten yang tergabung pada *smart city* studi kasus di Provinsi Jawa Barat. Terdapat enam parameter dalam pembobotan performansi *webiste* yang digunakan dalam penelitian ini yaitu *first-contentful-paint, first-meaningful-paint, speed-index, interactive, first-cpu-idle, max potential first input delay.* Nilai pembobotan tersebut dibandingkan dengan *dataset* Google, CrUX (*Chrome User Experience Report*). Hasil pengujian menunjukkan bahwa mayoritas *website* memiliki rata-rata performa 38,7 (*mobile*) 63,6 (*web*) dari 530 *website* yang tergabung dalam *smart city*, dengan hasil terbaik diperoleh Kota Bandung.

**Kata Kunci:** *Website Performance, Smart City, Jawa Barat, Pelayanan Publik,*

# PENDAHULUAN

Berdasar Survey (APJII, 2017) 44,16% pengguna internet Indonesia memilih perangkat mobile untuk mengakses internet, yang memilih komputer/laptop hanya 4,49%, dan menggunakan keduanya sebesar 39,28%. Berdasarkan (Mohorovičić, 2013) untuk mendapatkan pengalaman pengguna terbaik, suatu *website* perlu menyesuaikan ukuran layar dan resolusi dari perangkat/gawai. Karakter pengguna perangkat mobile cukup menarik dipandang dari hubungannya dengan kecepatan *load* suatu *website*. Dalam riset (Google, 2018) di Asia Tenggara dikatakan bahwa 53% pengguna mobile akan meninggalkan suatu *website* jika waktu *load* lebih dari 3 detik. Di sisi lain, sejak 2017 pemerintah melalui Kemenkominfo, tengah menggalakkan Program 100 Kota-Kabupaten *Smart City* (Kominfo, 2017). Gerakan tersebut bertujuan membimbing Kabupaten/Kota dalam menyusun *Masterplan Smart City* agar dapat lebih memaksimalkan pemanfaatan teknologi, baik dalam meningkatkan pelayanan masyarakat maupun mengakselerasikan potensi yang ada di masing-masing daerah. Kota/kabupaten yang terpilih menjadi calon *smart city* tentunya dilihat dari berbagai pertimbangan, salah satunya adalah akses jaringan. Di Indonesia sendiri, jangkauan jaringan 4G tergolong luas, dengan tingkat availability 83,5%, meski dengan kecepatan yang belum maksimal (Kompas, 2019). Dari data perilaku dan jumlah pengguna mobile yang besar tadi, seharusnya hal tersebut menjadi perhatian serius layanan *smart city*. Perlu adanya audit/uji guna memonitor kualitas *website* di instansi, kota/kabupaten apalagi yang sudah berjuluk kota pintar atau *smart city*. Gunanya tentu saja sebagai umpan balik atas pelayanan publik yang diberikan lewat teknologi *website*. Apalagi bila menilik data *website* INAPROC (Portal Pengadaan Nasional adalah pintu gerbang sistem informasi elektronik yang terkait dengan informasi Pengadaan Barang/Jasa secara nasional yang dibangun dan dikelola oleh Lembaga Kebijakan Pengadaan Barang/Jasa Pemerintah - Republik Indonesia) dengan kata kunci "*website*", pengadaan

*website* tergolong berbiaya tinggi, yaitu di atas Rp 50.000.000 (INAPROC, 2017), harusnya hal tersebut sebanding dengan kualitas/performa *website* yang dihasilkan.

Di antara uji performa yang dilakukan, yaitu untuk menjawab beberapa hal berikut: berapa skor/nilai rata-rata performa *website* yang ada di instansi, kota/kabupaten berjuluk *smart city* di Jawa Barat ? Dalam hal responsiveness *website*, bagaimana perbandingan skor performa mode mobile dibandingkan dengan mode desktop ? Mana daerah yang sudah baik skor performa *website*-nya ? Hal-hal tersebut menjadi acuan untuk perbaikan kualitas performa *website* terkait pelayanan terhadap publik, juga gambaran umum bagaimana kondisi kualitas *website* yang telah dibuat oleh tiap instansi, kota/kabupaten, dari tingkat provinsi, hingga kecamatan dan desa. Untuk mengukur performa *website* modern adalah feedback visual di saat memuat halaman *website*. Saat *browser* melakukan request kepada server, halaman kosong/blank page ditampilkan di layar. Kemudian pada titik tertentu, sesuatu yang lain muncul secara default untuk "menggambar" suatu elemen pada latar belakang tampilan, untuk momen pertama kalinya. Momen ini diistilahkan dengan *First Paint*. Selanjutnya, *browser* dapat melakukan *render*, seperti teks, webfont, gambar (termasuk latar belakang), konten berformat SVG, dan lain-lain. Momen ini diistilahkan dengan *First Contenful Paint*, saat itulah pengguna mulai dapat mengonsumsi konten halaman *website* (W3C, 2017).

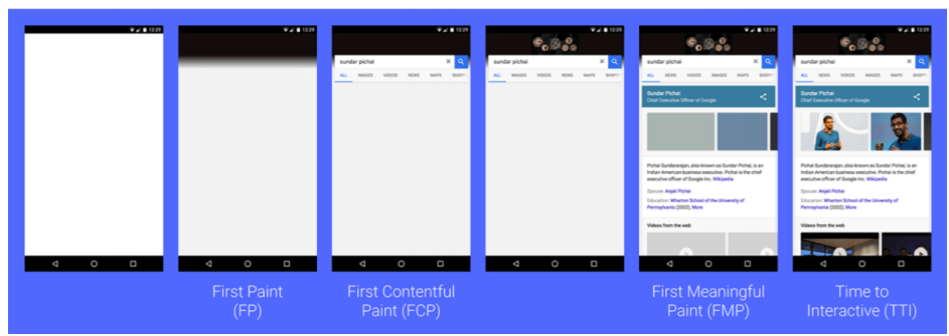

Gambar 1. *Screenshot* dari tahapan *load* suatu *website* berdasar waktu (Walton, 2018)

Suatu halaman *website* biasanya terdiri dari dokumen *Hypertext Markup Language* (HTML), file *Cascading Style Sheet* (CSS), dan *file* JavaScript (JS). Dokumen HTML mendefinisikan struktur halaman, CSS digunakan untuk memperindah tampilan, dan JS digunakan untuk fitur dan konten dinamis pada *website* (W3C, 2017).

**METODE PENELITIAN**

Dalam rangka mengiring dan memonitor kualitas *website* (dalam pelayanan masyarakat seputar informasi dan layanan publik), terkait waktu *load* dan *responsiveness*. Pada penelitian (Manhas, 2013) dikatakan bahwa ada beberapa hal yang mempengaruhi waktu *load website*, di antaranya: konten halaman, jenis *browser*, lokasi geografis pengguna, *bandwidth,* dan lain-lain. Pada (Budiman, 2018), parameter yang digunakan untuk menguji performa *website*: Time to First Byte, First paint time, First contentful paint time, DOM interactive time, DOM loaded time content, Onload time, total page size, dan jumlah *request*. Sedangkan penelitian ini menggunakan acuan tanpa melihat jumlah *request*, yaitu dengan metrik Lighthouse v5 (Google, 2019), di mana menilai waktu *load* berdasarkan metode pembobotan pada 6 parameter: *first-contentful-paint, first-meaningful-paint, speed-index, interactive, first-cpu-idle, max potential first input delay.* Sementara pada penelitian (Asrese, 2019) pengujian dilakukan dengan parameter tambahan kecepatan jaringan operator penyedia. Di dalam penelitian ini pengukuran disimulasikan dengan kecepatan akses jaringan 4G dengan asumsi tingkat cakupannya di Jawa Barat sudah baik.

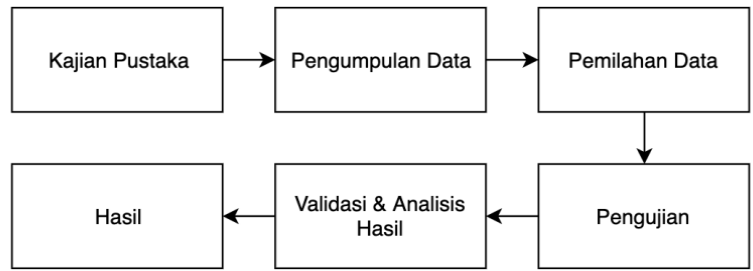

Gambar 2. Alur Penelitian

Dari kajian pustaka yang dilakukan, (Haapala, 2018) menguji performa *website* surat kabar di Finlandia, sementara (Jati, 2009) melakukan evaluasi kualitas *website e-government* dengan cakupan Asia (Singapura, Korea, Jepang, Hongkong, dan Malaysia), belum ada yang melakukan evaluasi yang sejenis (e-gov dan *smart city*) untuk kasus Jawa Barat, Indonesia. Dalam (Mohorovičić, 2013), didapatkan hasil bahwa untuk mendapatkan pengalaman pengguna terbaik, suatu *website* harus bersifat *responsive*, yang artinya menyesuaikan ukuran layar dan resolusi perangkat pengguna. Dari sana diterapkan audit performa *website* dengan cakupan *website* instansi dan kabupaten-kota berjuluk "*smart city*" di Provinsi Jawa Barat. Dilakukan pengumpulan data *website* yang ada di tingkat provinsi, kabupaten-kota, kecamatan, hingga desa. Dari data yang terkumpul (1.012 *website*) dipilih 530 *website* dari instansi dan kabupaten-kota berjuluk "*smart city*" saja berdasarkan (Kominfo, 2017) dan (Haryanto, 2018). Data didapatkan dari laman induk (Jabarprov.go.id), turunannya, dan pencarian manual. Data diuji dengan parameter dan pembobotan nilai berdasarkan Google PageSpeed Insight. Dari pengujian 530 *website* yang ada, hasilnya divalidasi secara manual, terutama yang nilainya mendekati batas atas 100 dan batas bawah 0. Hal ini dilakukan untuk menghindari adanya kesalahan uji. Kemudian dianalisis per daerah sebagai representasi performa *website* di instansi atau kota-kabupaten tersebut.

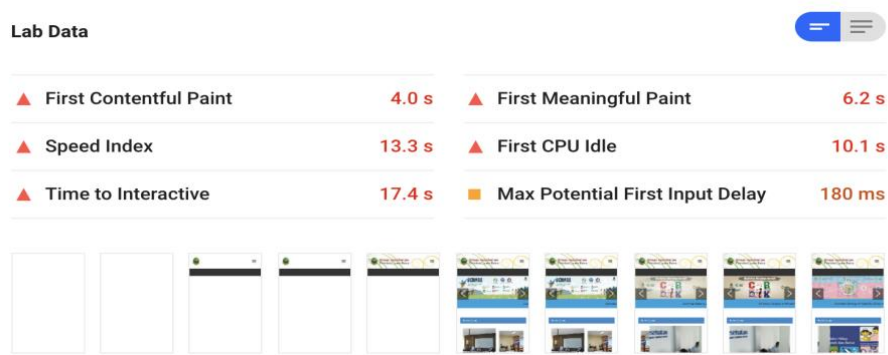

Gambar 3. Parameter skor performa *website* dan screenshot tahapan *load* suatu *website* (Google PageSpeed, 2019)

Kemudian skor dibandingkan dengan dataset Google, CrUX (Chrome User Experience Report) dalam meneliti pengalaman pengguna atas performa web. Dari uji tersebut, kemudian didapatkan skor untuk kemudian dikategorikan tingkat performa dan *rank*/urutannya.

Tabel 1. Contoh penghitungan dengan pembobotan berdasar penilaian Lighthouse v5

| Metric | Experience | Category Weighting | Performance Score |
|---|---|---|---|
| first-contentful-paint | Apakah muncul sesuatu? | 20.00% | 78 |
| first-meaningful-paint | Apakah ada yang berguna/bermakna? | 6.70% | |
| speed-index | Apakah ada yang berubah dari *blank*? | 26.70% | |
| interactive | Apakah bias digunakan/*usable*? | 33.30% | |
| first-cpu-idle | Permulaan konten dapat interaktif? | 13.30% | |
| max-potential-fid | Ada delay/jeda waktu? | 0.00% | |

Pembobotan diambil berdasarkan urgensi *experience* pengguna. Bobot terbesar diambil dari sisi *interactive* (33,3%) dengan pertimbangan sebagai gabungan dari waktu *load* dan waktu di mana pengguna mulai bisa berinteraksi dengan *website,* bukan sebatas muncul tampilan / *first-contentful-paint* (FCP) atau informasi / *first-meaningful-paint* (FMP). FCP mendapat bobot 20% karena cukup penting untuk membuat pengguna tidak meninggalkan *website*, disebabkan mulai muncul grafis dalam *load*-nya. FMP mendapat bobot 6,7% karena waktu *load* cukup lama bagi suatu halaman untuk menampilkan informasi yang bermakna. Sisanya pembobotan diambil dari *speed-index* dan potensi jeda/*delay.*

**HASIL DAN PEMBAHASAN**

Dari data kabupaten/kota yang sudah masuk dalam daftar Kominfo sebagai kota pintar/*smart city*, banyak di antaranya dari Provinsi Jawa Barat, tercatat ada 11 dari total 26 kabupaten/kota. Di antaranya: Kota Sukabumi, Kota Bandung, Kota Bekasi, Kota Bogor, Kota Cirebon, Kabupaten Cirebon, Kota Cimahi, Kabupaten Bogor, Kota Depok, Kabupaten Bandung, dan Kabupaten Indramayu.

Tabel 2. Daftar beberapa *website* institusi di Jawa Barat (jabarprov.go.id, 2019)

| No | Institusi | Tingkat | Laman |
|---|---|---|---|
| 1 | Biro Pemerintahan dan Kerjasama | provinsi | http://pemksm.jabarprov.go.id/ |
| 2 | Biro Hukum dan Hak Asasi Manusia | provinsi | http://jdih.jabarprov.go.id/ |
| 3 | Biro Pelayanan dan Pengembangan Sosial | provinsi | http://yanbangsos.jabarprov.go.id/ |
| 4 | Biro Badan Usaha Milik Daerah dan Investasi | provinsi | http://biroinvestbumd.jabarprov.go.id/ |
| 5 | Biro Perkenomoian | provinsi | http://biroperekonomian.jabarprov.go.id/ |

Provinsi yang memiliki visi "Jabar Juara" ini, ingin menjadi provinsi yang berbasis data dan teknologi (Jabar Digital Service, 2019). Pemanfaatan teknologi informasi di Jawa Barat mulai tampak dari tumbuhnya *website* milik instansi pemerintahan dan Satuan Kerja Perangkat Daerah (SKPD) hingga desa. Dari *website* tingkat provinsi hingga ke kabupaten/kota di Jawa Barat, berhasil dikumpulkan 1.012 *website*.

Dari 1.012 *website* yang terdata, diambil 530 saja karena yang diambil hanya yang masuk daftar kabupaten/kota *smart city*. Hasil dari pengujian dapat dilihat sebagai berikut.

Tabel 3. Daftar hasil uji performa *website* di Jawa Barat

| No | Daerah | Rata-rata Skor Mobile | Rata-rata Skor Web | Tanggal Uji |
|---|---|---|---|---|
| 1 | Web SKPD Provinsi | 26,98 | 47,51 | 25/08/19 |
| 2 | Kab. Bogor | 11,94 | 40,28 | 25/08/19 |
| 3 | Kab. Bandung | 53,10 | 72,21 | 25/08/19 |

| 4 | Kab. Indramayu | 48,42 | 76,77 | 25/08/19 |
|---|---|---|---|---|
| 5 | Kota Bogor | 32,83 | 76,33 | 25/08/19 |
| 6 | Kota Bandung | 84,61 | 98,68 | 25/08/19 |
| 7 | Kota Bekasi | 22,79 | 47,62 | 25/08/19 |
| 8 | Kota Cimahi | 65,20 | 79,70 | 25/08/19 |
| 9 | Kota Depok | 46,90 | 74,50 | 25/08/19 |
| 10 | Kota Sukabumi | 40,86 | 83,00 | 25/08/19 |
| 11 | Kota Cirebon | 29,62 | 64,46 | 25/08/19 |
| 12 | Kab. Cirebon | 1,40 | 1,81 | 25/08/19 |
| | Rata-rata total | 38,7 | 63,6 | 25/08/19 |

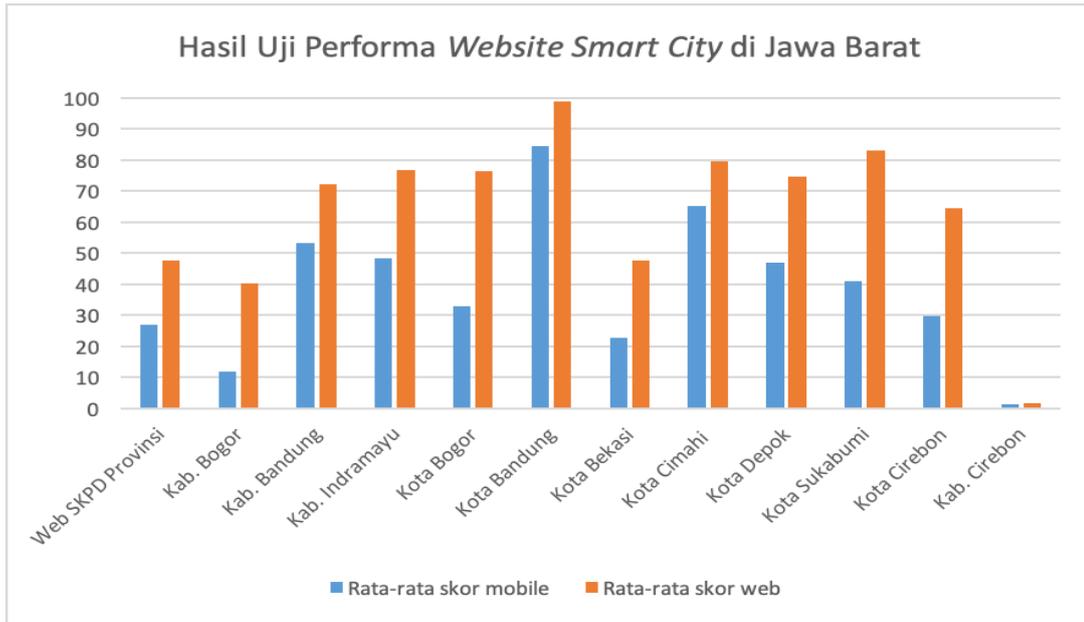

Gambar 4. Grafik skor web performance rata-rata per daerah

Hasil terbaik diperoleh Kota Bandung. Hal ini dapat dijadikan acuan untuk evaluasi pelayanan publik lewat *website* di Jawa Barat mulai tingkat provinsi hingga desa. Cakupan Jawa Barat dalam penilaian ini menjadi *pilot project* untuk diperluas lagi dan dengan jenis penilaian yang lebih beragam.

**SIMPULAN**

Dari 1.012 *website* yang terdata, diambil 530 saja karena yang diambil hanya yang masuk daftar kabupaten/kota *smart city*. Hasil dari pengujian menunjukkan bahwa mayoritas *website* memiliki rata-rata skor performa 38,7 (mobile) dan 63,6 (web), dengan hasil terbaik diperoleh Kota Bandung. Hal ini dapat dijadikan acuan untuk evaluasi pelayanan publik lewat *website* di Jawa Barat mulai tingkat provinsi hingga desa. Cakupan Jawa Barat dalam penilaian ini menjadi pilot project untuk diperluas lagi dan dengan jenis penilaian yang lebih beragam.

# DAFTAR PUSTAKA